\documentclass[citeautoscript,floatfix,aps,prl,onecolumn,superscriptaddress]{revtex4-1}
\usepackage{graphicx}
\usepackage{amsmath}
\usepackage{amssymb}
\usepackage{dcolumn}
\usepackage[version=4]{mhchem}
\usepackage{latexsym}
\usepackage{rotating}
\usepackage{epstopdf}
\usepackage[usenames,dvipsnames]{xcolor}
\usepackage{float}
\usepackage{soul}
\usepackage{epsfig}
\usepackage{psfrag}
\usepackage{natbib}
\usepackage{bm}
\usepackage{eucal}
\usepackage{mathrsfs}
\usepackage{braket}
\usepackage{enumerate}
\usepackage{longtable}
\usepackage{bm}
\usepackage{hyperref}
\usepackage{amsfonts}
\usepackage{dcolumn}
\usepackage{bm}
\usepackage{changes}

\newcommand{\be}{\begin{equation}}
\newcommand{\ee}{\end{equation}}
\newcommand{\bn}{\begin{eqnarray}}
\newcommand{\en}{\end{eqnarray}}

\def\x2y2{{x^2-y^2}}

\usepackage{color} 

\usepackage{hyperref}
\hypersetup{
colorlinks=true,final=true,
        linkcolor=red,
        citecolor=blue,
        filecolor=blue,
        urlcolor=blue,
}
\begin{document}

\title{First principles vs second principles: Role of charge self-consistency in strongly correlated systems}
\author{Swagata Acharya}
\affiliation{Institute for Molecules and Materials, Radboud University, {NL-}6525 AJ Nijmegen, The Netherlands}	
\email{swagata.acharya@ru.nl}
\author{Dimitar Pashov}
\affiliation{ King's College London, Theory and Simulation of Condensed Matter,
	The Strand, WC2R 2LS London, UK}
\author{Alexander N. Rudenko}
\affiliation{Institute for Molecules and Materials, Radboud University, {NL-}6525 AJ Nijmegen, The Netherlands}
\author{Malte R\"{o}sner}	
\affiliation{Institute for Molecules and Materials, Radboud University, {NL-}6525 AJ Nijmegen, The Netherlands}
\author{Mark van Schilfgaarde}
\affiliation{ King's College London, Theory and Simulation of Condensed Matter,
	The Strand, WC2R 2LS London, UK}
\affiliation{National Renewable Energy Laboratory, Golden, CO 80401, USA}	
\author{Mikhail I. Katsnelson}
\affiliation{Institute for Molecules and Materials, Radboud University, {NL-}6525 AJ Nijmegen, The Netherlands}


\begin{abstract}
First principles approaches have been successful in solving many-body Hamiltonians for real materials to an extent when
correlations are weak or moderate. As the electronic correlations become stronger often embedding methods based on first
principles approaches are used to better treat the correlations by solving a suitably chosen many-body Hamiltonian with
a higher level theory. Such combined methods are often referred to as second principles approaches. At such level of the theory the self energy, i.e. the functional that embodies the stronger electronic correlations, is either a function of energy or momentum or both. 
The success of such theories is commonly measured by the quality of the self energy functional. However,
self-consistency in the self-energy should, in principle, also change the real space charge distribution in a
correlated material and be able to modify the electronic eigenfunctions, which is often undermined in second principles approaches. 
Here we study the impact of charge self-consistency within two example cases: TiSe$_{2}$, a three-dimensional charge-density-wave candidate material, and CrBr$_{3}$, a two-dimensional ferromagnet, and show how real space charge re-distribution due to correlation effects taken into account within a first principles Green's function based many-body perturbative approach is key in driving qualitative changes to the final electronic structure of these materials.
\end{abstract}
\maketitle

Density functional theory \cite{HK0,KS0,JGrev} has been the workhorse for material specific electronic structure calculations for the last half of the century. Despite enormous success in many respects, it has however some intrinsic limitations. 
First of all, although the Hohenberg-Kohn theorem \cite{HK0} guarantees the existence of {\it some} density
functional providing an exact ground state energy at a given charge density distribution $\rho$, its exact form is unknown. In
practice, this functional is considered as being local or almost local (generalized gradient corrections), which is
generally speaking an uncontrollable approximation (for detailed discussions see the review \cite{JGrev}). Next, and even
more importantly, the Kohn-Sham quasiparticles \cite{KS0} are, generally speaking, just auxiliary quantities to
calculate the total energy and their direct comparison with experimental spectroscopic information is hardly
justifiable. Although this is regularly done with partial excellent agreement, there are numerous counterexamples starting
from the famous ``gap problem'' in semiconductors \cite{gw}.

An alternative approach is based on the concept of Green's function functionals. Luttinger-Ward \cite{lward} and
Baym-Kadanoff \cite{BK} theorems respectively prove the existence of such functionals in- and out-of-equilibrium. 
Conceptually, this way is more attractive since the knowledge of an exact single- and two-particle Green's
functions guarantees an accurate description of spectroscopic properties of solids \cite{nozieres}. On the other hand,
again, an exact form of this functional is practically unknown and we have just its formal definition in terms of infinite
sums of skeleton free-leg diagrams \cite{lward,BK}. If we are interested in a description of subtle phenomena such as, e.g., the Kondo effect \cite{hewson} or nonquasiparticle states in half-metallic ferromagnets \cite{HMFMRMP}, the
necessary sequence of diagrams seems to be too complicated to be practically taken into account for a complete
first-principle realization.

Therefore, alternative embedding approaches were introduced which combine first-principle calculations with model treatments to describe the strong correlations within some low-energy subspace. This way, weakly correlated states at high energies are  described within a low-level theory, while the strongly correlated sub-space is treated in  higher level approaches. This is popularly done by mapping the low-energy space to multi-band generalized Hubbard models, which are afterwards often solved, e.g., using dynamical mean-field theory (DMFT) \cite{ldadmft1}, a program suggested and called ``LDA$^{++}$'' in Ref.~\onlinecite{lda++} and which we refer to in the following as ``second principles''.
In many cases, this leads to a dramatic improvement of description of strong correlation effects in real materials with itinerant-electron magnets \cite{ldadmft2} and heavy-fermion compounds \cite{HauleHF} being two major successful  examples (for detailed reviews see \cite{HeldReview,kotliarDMFTRMP,HMFMRMP}). Of course, DMFT \cite{georgesDMFTRMP} is a {\it local} approximation which takes only the energy dependence of electronic self energy into account and completely neglects its momentum dependence. However, the latter can be taken into account via various beyond-DMFT diagrammatic approaches \cite{beyondDMFTReview}, rendering it a technical problem rather than a fundamental one. Also the way how one can map the first-principle electronic structure onto efficient Hamiltonians can be, in principle, improved. The contemporary way is based on the so-called constrained RPA (cRPA) approach \cite{cRPA,honer} but there are no principle obstacles to improve it further if necessary. 

A key impediment to second principles approaches is, however, that multiple energy scales are operative: the high-energy scales controlling low-energy fluctuations cannot be integrated out without model assumptions.  
High-energy scales contain information about chemistry and disorder specific to real materials.  
Yet while first principles theories contain this information, their application to strongly correlated systems has been limited. 
In weakly correlated materials, ``first principles'' approaches tend to predominate because they rely on a minimum of model assumptions, and are often predictive.  
This is not the case when correlations are strong because standard methods, usually based on extensions of density functional theory (DFT), lack the sophistication to encapsulate the strong spin and charge fluctuations, or the fidelity to characterize one-particle properties near the Fermi level (which are essential to capture low-energy excitations characteristic of correlated systems); nor are they adequately equipped to generate the (two-particle) suceptibilities.  
Even in weakly correlated cases, dynamical screening (perhaps the most important many-body effect~\cite{Martinbook}) is not well treated by such standard one-body descriptions~\cite{Neaton06,Bruneval06b}.  
The difficulties are even more severe for spin fluctuations, where the characteristic energy scale for excitations can be very small.

Models such as the Hubbard Hamiltonian do indeed contain, within some region of parameter space, key many-body effects such as the metal-insulator transition, pseudogap phases, quantum criticality, and both conventional and unconventional superconductivity.  
Thus, applications of this model has become the canonical approach to characterizing such phenomena. 
However, the limits to such an approach become apparent when the high-energy scales that control parameters for the low-energy ones are nontrivial. 
Furthermore, within any model Hamiltonian, there are only two possibilities for the correlation effects to  modify the electronic structure, namely via the energy- and/or the momentum-dependencies of the self-energy $\Sigma$. 
In first-principle approaches there is, however, an additional possibility in form of the charge self-consistency. That is, if the correlation effects can essentially modify the charge distribution $\rho$, than one needs to recalculate the model Hamiltonian at every iteration which makes the mapping procedure very cumbersome or even practically useless. The physical question of fundamental importance is the following: can  a real-space charge redistribution due to correlation effects be {\it qualitatively} important leading not just to a moderate renormalization of the model parameters but also to a reconstruction of the electronic structure beyond {\it any} purely model consideration? In this work we give a positive answer on this question providing two examples, namely, TiSe$_2$ and CrBr$_3$. 

	\begin{figure}[tbp]
		\begin{center}
			\includegraphics[width=0.45\columnwidth]{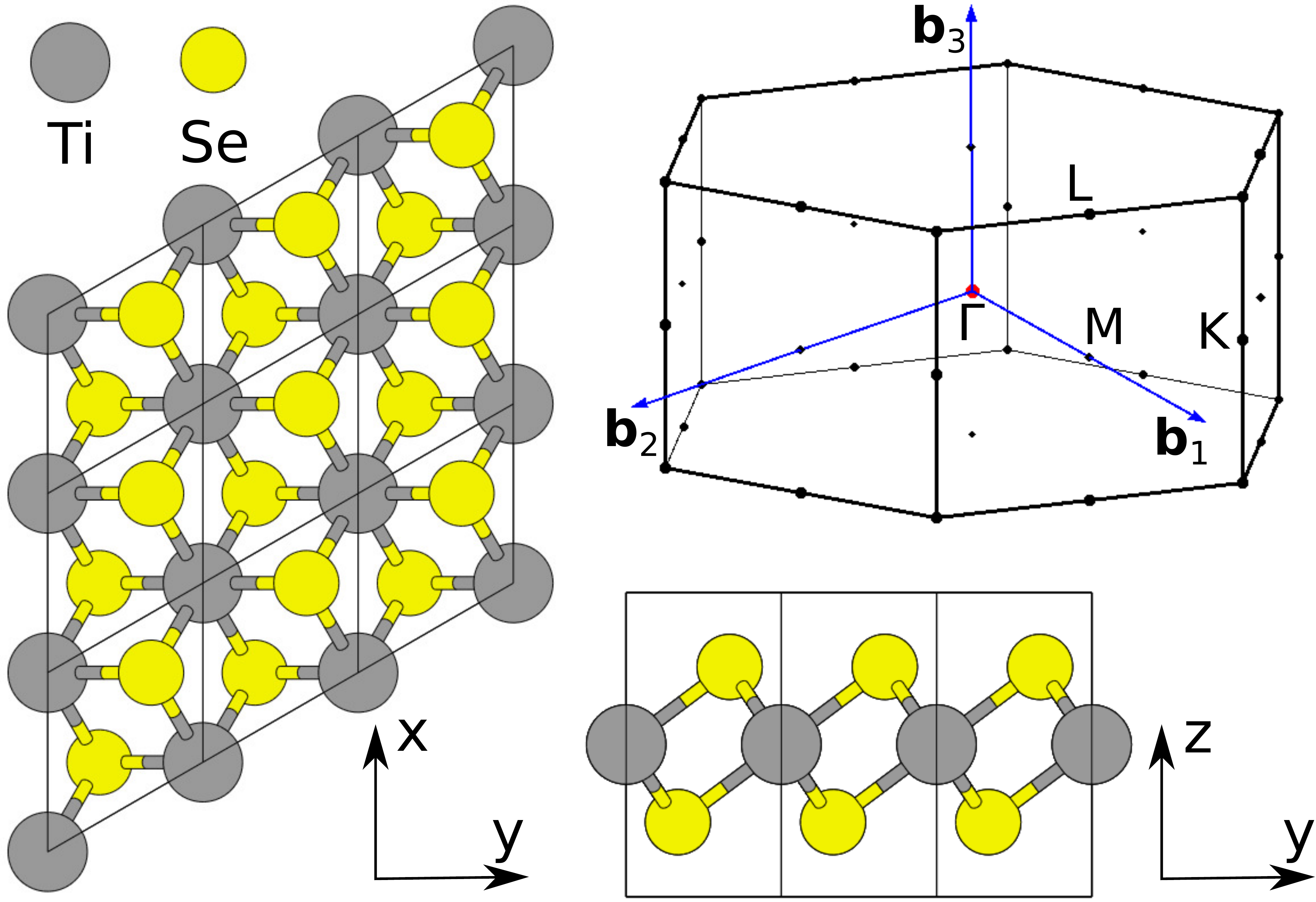}
			\quad \quad \quad \quad
			\includegraphics[width=0.45\columnwidth]{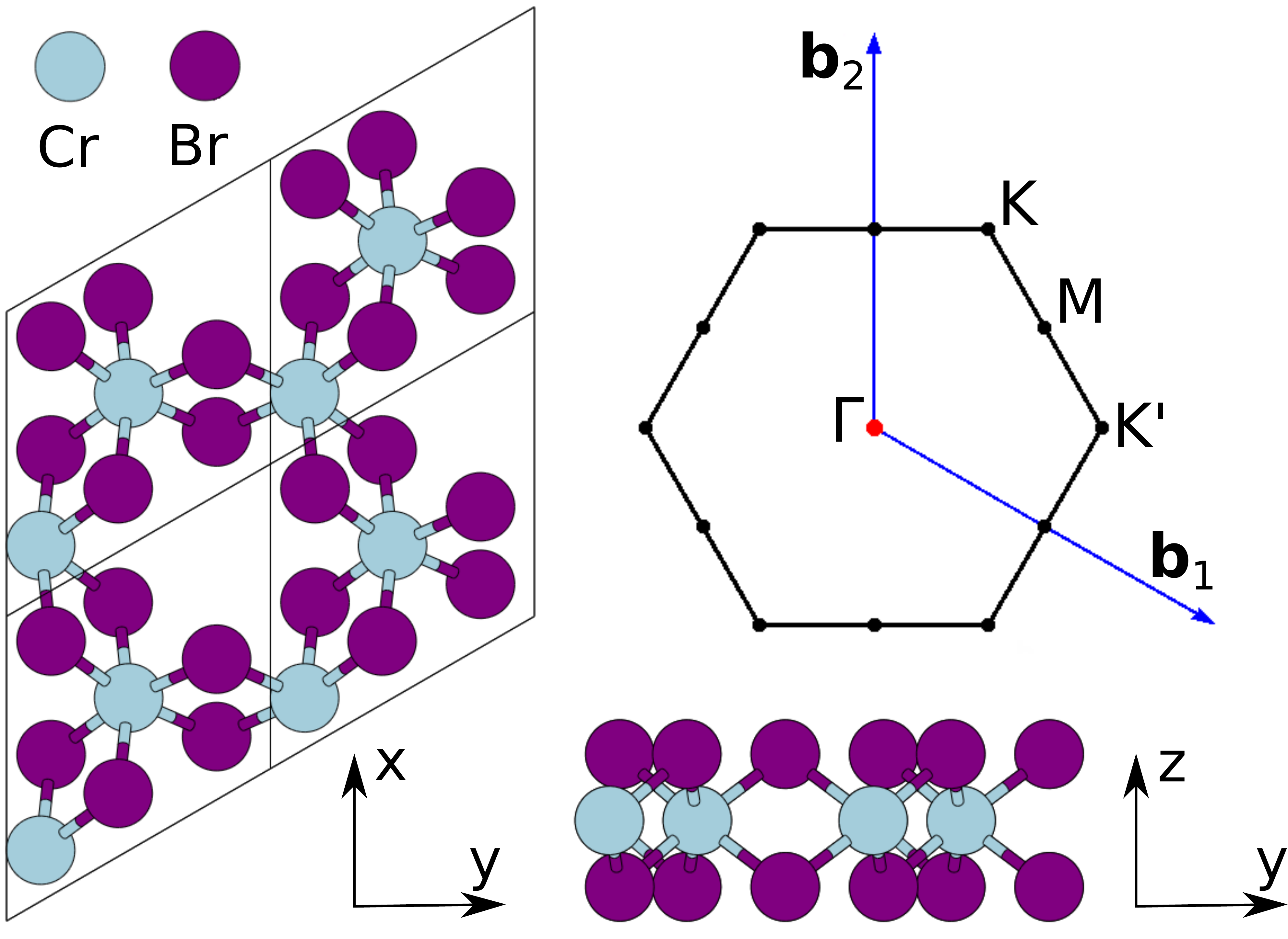}
			\caption{Ball-and-stick model of the crystal structure, and the Brillouin zone of TiSe$_2$ (left) and 1L-CrBr$_3$ (right). ${\bf b}_1$, ${\bf b}_2$, and ${\bf b}_3$ denote reciprocal lattice vectors. 
}
			\label{fig:struc}
		\end{center}
	\end{figure}

We show in the following how different levels of theory significantly modify the effective one-body potential through changes in the electron density.  
To this end, we employ three different levels of theory: 
the local-density approximation (LDA),
quasiparticle self-consistent \emph{GW} theory~\cite{mark06qsgw,Kotani07,questaal_paper} (QS\emph{GW}), which, in contrast to conventional \emph{GW},
modifies the charge density and is determined by a variational principle~\cite{Beigi17}, 
and finally an extension of QS\emph{GW}, where the polarizability needed to construct \emph{W} is computed including vertex corrections (ladder diagrams) by solving a Bethe-Salpeter equation (BSE) for the two-particle Hamiltonian~\cite{cunningham21}.  
We denote the latter QS$G\widehat{W}$, with the substitution $W{\rightarrow}\widehat{W}$ signifying that a BSE was solved to compute \emph{W}.  
 These first-principles approaches allow us to carefully analyze the impact of the full charge self consistency taking correlation effects with increasing diagrammatic precision into account.

In terms of diagram classes taken into account QS\emph{GW} and QS$G\widehat{W}$ represent the forefront of currently available first-principle approaches.
As we show, it is essential that the first-principles starting point is of sufficiently high fidelity to capture physics the second-principles scheme cannot reach. First-principles schemes are too cumbersome to handle more than a limited class of diagrams, and it may still be true in general that second-principles schemes may still be needed to capture physics outside the reach of the first-principles scheme. Low-energy spin fluctuations, Kondo effect and non-quasi-particle states in weakly doped Mott insulators and half-metallic ferromagnets seem to be the archetypal example of this. For TiSe\textsubscript{2} and CrBr\textsubscript{3}, QS\emph{GW} / QS$G\widehat{W}$ appears, however, to adequately describe most physical observables, obviating the need for second-principles schemes.

\noindent\emph{\bf TiSe$_{2}$:}

TiSe$_{2}$ is a layered diselenide compound with space group $P\bar{3}m1$ (Fig.~\ref{fig:struc}).  Below 200K, it undergoes a phase transition
to a charge-density wave (CDW), forming a commensurate $2{\times}2{\times}2$ superlattice ($P\bar{3}c1$) of the original
structure.  At the transition there is a softening of the zone boundary phonon accompanied by changes in the transport properties
\cite{DiSalvo76, Holt01}.

A number of works have tried to determine the energy dispersion around the Fermi level in both phases with a special focus on the
overlap/gap between the Se-4\emph{p} valence band  at $\Gamma$ and the Ti-3\emph{d} conduction band at L, sometimes with
discordant results.  In the high-temperature $P\bar{3}m1$ phase, reports range from predicting a semimetal (overlap
$<$120\,meV) between these bands, to an insulator with $<$60\,meV gap, depending on the
study~\cite{Rossnagel02,Anderson85,Traum78,Stoffel85,Pillo00,Kidd02,Rasch08,Cava07}.  In the CDW $P\bar{3}c1$ phase there is a greater consensus, namely that the gap
is small and positive.  In brief, upon cooling the system, the CDW transition induces a distortion that either
(slightly) increases the existing gap or leads to a gap opening between these bands with the gap for the
CDW phase being $\sim$100-150\,meV~\cite{Stoffel85,Pillo00,Rossnagel02,Kidd02,Cava07}.

At high temperature, the positive or negative (i.e., overlap) indirect gap between Se-4\emph{p} and Ti-3\emph{d} bands
is larger than the negative gap obtained with standard DFT calculations. \added{In fact,} DFT is not helpful because it predicts a
negative gap in both the undistorted and CDW phases.  Bianco \emph{et al.}~\cite{Mauri15} finds with LDA+\emph{U} (\emph{U}=3.9
eV) a gap of 14\,meV in the $P\bar{3}m1$ phase and 200\,meV in the CDW phase.  LDA+$U$ is a kind of ``second
principles'' method because the answer depends $U$, which is not known.  To check whether the negative gap is merely an
artifact of the model, Cazzaniga \emph{et al.} \cite{Cazzaniga12} considered a $G_0W_0$ calculation based on the LDA, and found a
gap of $\sim$200\,meV in the high-temperature $P\bar{3}m1$ structure.

We show here that while $GW$ does indeed modify the quasiparticle spectrum, the true situation is more complex.  This is
because not only the eigenvalues but the density is significantly renormalized relative to the LDA.
This induces a corresponding change in the effective potential through the inverse of the susceptibility,
$\chi^{-1}(x_1,x_2) = \delta V(x_1)/\delta n(x_2)$.  What appears to be special about TiSe$_{2}$ is that
$\chi^{-1}(1,2)$ is large, and the correction to the LDA density modifies the effective potential $V$ in such a way as
to \emph{reduce} the splitting between occupied and unoccupied levels.  This is very unusual: it has long been
established that in the vast majority of cases, \emph{GW} based on the LDA continues to underestimate the gap in
semiconductors, albeit less so than the LDA~\cite{mark06adeq}.

\begin{figure}
	\begin{center}
\includegraphics[width=0.7\columnwidth]{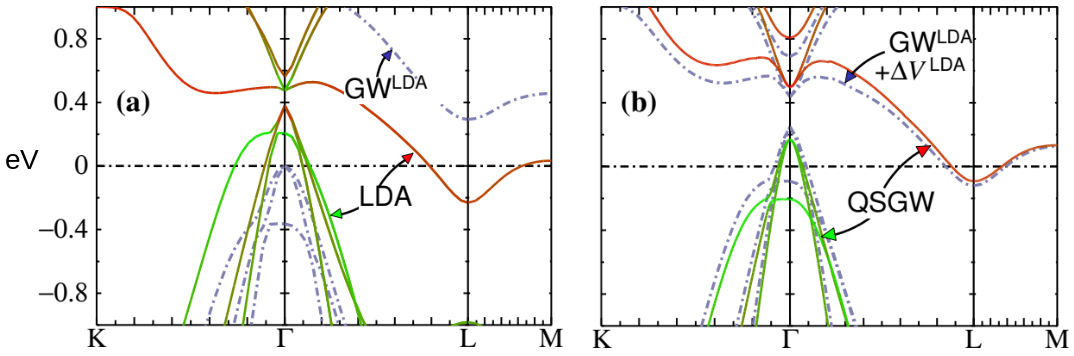}
\caption{Energy bands of the undistorted $P\bar{3}c1$ structure.  $(a)$: solid lines are LDA results, with red and green
  depicting a projection onto Ti and Se orbital character, respectively.  Blue dashed line shows shifts calculated in
  the $GW$ approximation based on the LDA, as described in the text. $(b)$: blue dashed line shows results from $GW$
  based on the LDA (same $\Sigma$ as in panel $(a)$), with an extra potential $\Delta V^\mathrm{LDA}$ deriving from a
  $\rho$ shift computed from the rotation of the LDA eigenvectors.  Solid lines are QS\emph{GW} results, with
  the same color scheme as in panel $(a)$.  }
\label{fig:fig1}
	\end{center}
\end{figure}

These effects can only be found through self-consistency.  QS\emph{GW} is ideally suited for this case, as its
excitation spectra are generally superior to fully self-consistent
$GW$~\cite{holm98,Shirley96,BelashchenkoLocalGW,Tamme99,Setten16}.
We find that the $P\bar{3}m1$ is indeed semimetallic, as is the case with DFT, but for different reasons.
We first revisit the \emph{GW} calculation of the undistorted $P\bar{3}c1$ structure, but with some modifications:\\
$\bullet$ we did not include a $Z$ factor.  There are various justifications for this, most notably as an approximate way to
incorporate self-consistency in $G$ with fixed $W$; see Appendix in Ref.~\onlinecite{mark06adeq}.
Omission of $Z$ tends to widen bandgaps.\\
$\bullet$ the full matrix $G^\mathrm{LDA}W^\mathrm{LDA}$ is used, in the QS\emph{GW} sense \cite{Kotani07}:
\begin{eqnarray}
\Sigma^0_{ij} = \frac{1}{2}\sum_{ij} |\psi_i\rangle
       \left\{ {{\rm Re}[\Sigma(\varepsilon_i)]_{ij}+{\rm Re}[\Sigma(\varepsilon_j)]_{ij}} \right\}
       \langle\psi_j|.
\label{eq:veff}
\end{eqnarray}

Panel $(a)$ of Fig.~\ref{fig:fig1} shows LDA and $G^\mathrm{LDA}W^\mathrm{LDA}$ bands similar to the
$G^\mathrm{LDA}W^\mathrm{LDA}$ calculation of Ref.~\onlinecite{Cazzaniga12}.  Focusing on the LDA bands, the highest occupied
state at $\Gamma$ turns red very close to $\Gamma$, indicating the penetration of the Ti-derived conduction band into the
valence band (resembling a ``negative'' bandgap).  This is an artifact of the LDA's well known tendency to underestimate
splittings between occupied and unoccupied states, and as standard $G^\mathrm{LDA}W^\mathrm{LDA}$ increases this
separation (blue dashed lines).  The (indirect) $G^\mathrm{LDA}W^\mathrm{LDA}$ gap of 300\,meV is slightly larger than
Ref.~\onlinecite{Cazzaniga12}, in line with the unit $Z$ factor used in the present calculation.

Fig.~\ref{fig:fig1}$(b)$ shows that self-consistency is crucially important in TiSe$_{2}$.  The off-diagonal elements of
$\Sigma^0_{ij}$ modify the density $n(\mathbf{r})$ and thus $V(\mathbf{r})$.  A simple way to estimate $\Delta V$ is to
make an ansatz that the LDA adequately yields $\chi^{-1}=\delta V/\delta n$.  For a modified $\bar{n}$ the potential
becomes $V(\bar{n}) = \Sigma^0-V_\mathrm{xc}(n^\mathrm{LDA})+V_\mathrm{xc}(\bar{n})$.  $\bar{n}$ can be determined
self-consistently in the usual manner by adding a fixed external potential $\Sigma^0-V_\mathrm{xc}(n^\mathrm{LDA})$ to
the LDA Hamiltonian and allowing it to go self-consistent.  Remarkably, the gap becomes negative again, as shown by the
blue dashed lines in Fig.~\ref{fig:fig1}$(b)$, but the dispersion is very different from the LDA.  In particular the
inverted gap character at $\Gamma$ disappears, which is topologically essential for a gap to form at $\Gamma$.  The
quality of the ansatz can be checked by carrying out a complete QS\emph{GW} calculation.  This is shown as solid lines in
Fig.~\ref{fig:fig1}$(b)$, and it demonstrates the ansatz is reasonable.  As we will show elsewhere,
the observed low-temperature gap forms as a consequence of the charge density-wave instability.

\noindent\emph{\bf CrBr$_{3}$:}

Monolayer (1L) of CrBr$_{3}$ is a two-dimensional ferromagnetic (FM) insulator where the magnetic moments of monolayer
CrBr$_{3}$ align normal to the plane (see Fig.~\ref{fig:struc} for the crysal structure).  The spontaneous magnetization persists in monolayer CrBr$_{3}$ with a Curie
temperature of 34 K~\cite{zhang}. Within a purely atomic picture, fully determined by the crystal field environment and
the Hund's multiplet structure~\cite{molina}, the low energy properties of the materials and the magnetism should be
entirely governed by Cr-\emph{d} electrons. However, the ligands, their masses and the number of core states they have,
play an integral role in determining the low energy properties of CrBr$_{3}$. In a separate work we discuss the role of
the ligands like Cl, Br and I in determining the crucial low energy properties of the entire class of 1L Chromium
trihalides~\cite{swagcrx1}. The Br-\emph{p} states strongly hybridize with Cr-\emph{d} states in CrBr$_{3}$. In the
present work we show how charge self-consistency at different levels of the theory controls the nature of the
eigenfunctions and the Br component in the valence band manifold of 1L CrBr$_{3}$.
 
\begin{figure}
	\begin{center}
	\includegraphics[width=0.7\columnwidth]{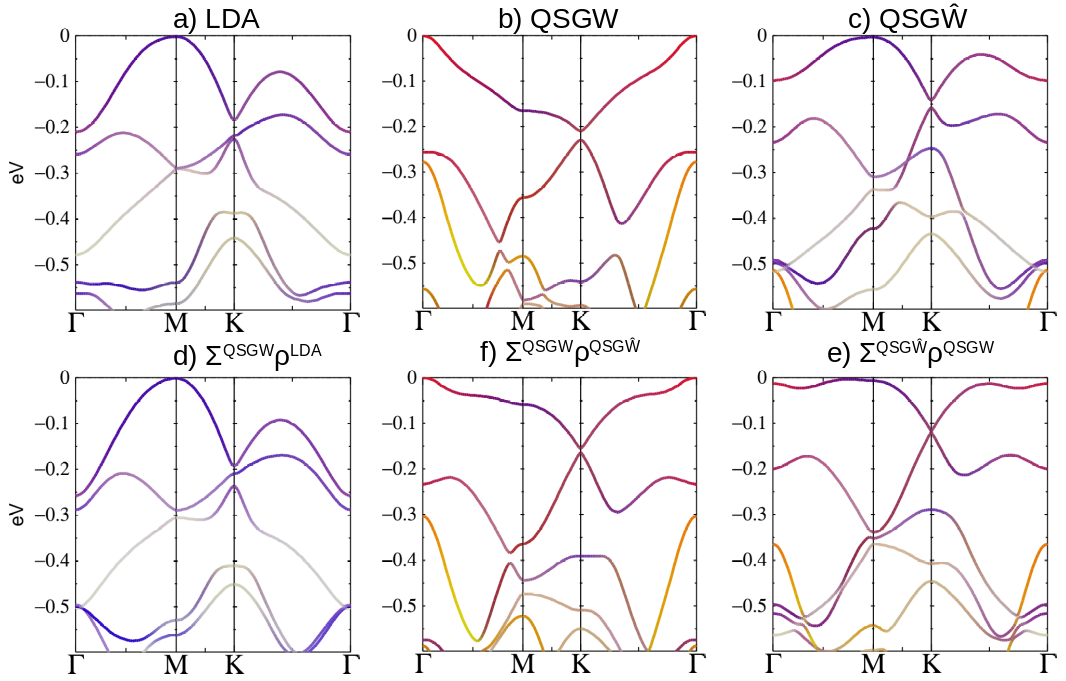} 
		\caption{{\bf Electronic
		Band Structure of CrBr$_{3}$} : (a) LDA, (b) QS\emph{GW} and (c) QS$G\widehat{W}$. 
		$\widehat{W}$, we mean the polarizability that enters into $W$ is calculated beyond the time-dependent
		Hartree approximation (RPA) but also includes ladder diagrams computed from a four-point Bethe-Salpeter
		equation (BSE).  This significantly improves $W$ as seen by comparing the macroscopic dielectric
		function to experiment.   Colors correspond to Br-$p_{x}{+}p_{y}$
		(red), Br-$p_{z}$ (green), Cr-\emph{d} (blue).  A fourth color (white) selects the Cr-$e_g$ manifold, 
                and washes out the color to the extent it is present.  Topmost valence bands of LDA and BSE
		have similar shape, apart from a strong narrowing the bandwidth. The shape changes in QS\emph{GW}.
                with the VBM shifting to $\Gamma$.  (d) $\Sigma^{\mathrm{QS}GW}[\rho(\mathrm{LDA})]$, (e)
                $\Sigma^{\mathrm{QS}GW}[\rho(\mathrm{{QS}G\widehat{W}})]$, and (f) $\Sigma^{\mathrm{QS}G\widehat{W}}[\rho(\mathrm{QS}GW)]$.}
		\label{fig:band}
	\end{center}
\end{figure}

We simulate the free standing 1L of ferromagnetic CrBr$_{3}$ within LDA, QS$GW$ and QS$G\widehat{W}$.  We also perform a
rigorous check for vacuum correction to all relevant observables by increasing the vacuum size from 20 \AA\ to 80
\AA~\cite{swagcrx1}. We check for convergence and scaling of band gap and the dielectric constant $\epsilon_{\infty}$
with vacuum size as discussed in a separate work~\cite{swagcrx1}. We observe that FM-1L CrBr$_{3}$ is an insulator with
1.3 eV of electronic band gap in LDA, which is significantly lower than in QS\emph{GW} yielding a gap of 5.7 eV. The large enhancement in QS\emph{GW} band gaps relative to
the LDA is standard in polar compounds~\cite{mark06qsgw}.  Nevertheless, within the random phase approximation (RPA), it
has long been known that $W$ is universally too large~\cite{albrecht,Rohlfing98b}, which is reflected in an
underestimate of the static dielectric constant $\epsilon_\infty$.  Empirically, $\epsilon_\infty$ seems to be
underestimated in QS\emph{GW} by a nearly universal factor of 0.8~\cite{deguchi}, for a wide range of
insulators~\cite{chantis06a,bhandari18} resulting in slightly overestimated~\cite{mark06qsgw} band gaps.  This can be
corrected by extending the RPA screening to introduce an electron-hole attraction in virtual excitations.  These extra (ladder)
diagrams are solved by a BSE, and they significantly improves on the optics, largely eliminating the discrepancy in
$\epsilon_\infty$ \cite{Cunningham18}.  When ladder diagrams are also added to improve $W$ in the \emph{GW} cycle
($W{\rightarrow}\widehat{W}$), it significantly improves the one-particle gap as well~\cite{Kutepov16,cunningham21}. In detail, our QS$G\widehat{W}$ implementation is self-consistent in the sense that the updated $\widehat{W}$ also
updates $\Sigma$ and hence the cycle continues until $\widehat{W}$, $\Sigma$ and $G$ converge iteratively. This scenario is played out in \ce{CrBr3}: the QS\emph{GW} bandgap is slightly larger than QS$G\widehat{W}$ bandgap, as
seen in Table~\ref{table1}.  
Also we converge the
observables like band-gap and $\epsilon_{\infty}$ by increasing the size of the two-particle Hamiltonian within our
self-consistent BSE implementation. We find that for CrBr$_{3}$ to converge both observables we find it necessary to
include 24 valence bands and 24 conduction bands in two-particle Hamiltonian that we solve within
BSE. Larger sizes of the two-particle Hamiltonian did not lead to any changes in these observables. The convergence with respect to the number of states entering into the two-particle
is significantly slower than in simple \emph{sp} semiconductors. Once the
two-particle Hamiltonian size is converged, we converge the observables in terms of vacuum size.  

Next we examine independent variations of the Hartree potential, via the density $\rho$, and the self-energy $\Sigma^0$. In the \emph{GW} case, $\Sigma^0$ denotes the quasiparticlized version of the dynamical self-energy $\Sigma(\omega)$;
for the LDA it denotes the LDA exchange-correlation potential. Unless stated otherwise, results are presented with
$\rho$ and $\Sigma^0$ internally self-consistent.  Considering this case at first, there is a remarkable difference
between the LDA and QS\emph{GW} electronic band structures.  Within LDA (see Fig.~\ref{fig:band}(a)), the valence band maximum falls at the M point, while within QS\emph{GW} it shifts to the $\Gamma$ point (see Fig.~\ref{fig:band}(b)).
The eigenfunctions are also quite different: the Br contribution to the low energy valence band manifold is
significantly larger within QS\emph{GW}.  At a still higher level of theory replacing $W{\rightarrow}\widehat{W}$, a
portion of the strong perturbation of the LDA band structure is partially undone (Fig.~\ref{fig:band}(c)); shifts Br
contribution to the valence eigenfunctions in the direction of the LDA (see table~\ref{table1}). This is
readily understood as a softening of $W$ by the ladder diagrams, as noted above.  With the QS$G\widehat{W}$, the bandgap in
CrBr$_{3}$ is reduced slightly to 4.65 eV.  
The top most valence band in QS$G\widehat{W}$ has a shape similar to LDA but the band gap is approx. three times as large and the valence bandwidth gets renormalized by a factor of $\sim 2$. The observation that QS$G\widehat{W}$
more closely resembles the LDA than QS$G{W}$ is remarkable and calls for further analysis.

\begin{table}
	\footnotesize
	\begin{tabular}{ccccccccccc}
		\hline
		\hline
		variants & LDA & QS\emph{GW} & QS$G\widehat{W}$ & $\Sigma^{\mathrm{QS}GW}[\rho(\mathrm{LDA})]$ & $\Sigma^{\mathrm{QS}GW}[\rho(QSG\widehat{W})]$ & $\Sigma^{\mathrm{QS}G\widehat{W}}[\rho(\mathrm{QS}GW)]$   \\
		\hline
	$\%$ of Br &	31  & 69 & 37 & 23 & 58 & 47 \\     
	Cr-\emph{d} & 	4.44  & 4.3 &  4.35 & 4.36 & 4.32 & 4.31 \\
	gap (eV) & 1.3 & 5.7 & 4.65 & 6.0 & 5.7 & 4.69   \\		
	\hline
	\end{tabular}
	\caption{Fraction of spectral weight that the Halogen (Br) contributes to the total 
	within an energy window of 0 (Fermi energy) to 0.6 eV below the Fermi energy (bound states). The Cr-\emph{d}
	occupancies, and the electronic band gaps with different choices of self consistent $\rho$ and $\Sigma$ are
	shown.}
	\label{table1}
\end{table}

To this end, we consider independent variations of $\rho$ and $\Sigma^0$ in the following senses:

\noindent$\bullet$ $\rho$ from LDA and $\Sigma^0$ from QS\emph{GW}, which we denote as
$\Sigma^{\mathrm{QS}GW}[\rho(\mathrm{LDA})]$.  This scheme produces a valence band structure similar to LDA (see Fig.~\ref{fig:band}(d)), but with 6.0\,eV electronic gap, close to the QS\emph{GW} gap of 5.7\,eV. This clearly establishes the important role of the density in determining the effective
one-body hamiltonian.  As in the case of simple \emph{sp} tetrahedral semiconductors where the LDA density is already
rather good, the gap change is mainly controlled by the nonlocality in the self-energy which the LDA
misses~\cite{Kotani98,Gruning06}.

\noindent$\bullet$ $\rho$ from QS$G\widehat{W}$ and $\Sigma$ from QS\emph{GW}, which we denote as
$\Sigma^{\mathrm{QS}GW}[\rho(QSG\widehat{W})]$ (see Fig.~\ref{fig:band}(e)).  Now the valence band
structure is much closer to the QS\emph{GW} band structure, although the top most valence band is significantly
narrowed.  Also, the gap is similar to QS\emph{GW} gap (5.7\,eV).  This tells us that the Hartree and many-body
contributions cannot be decoupled.  The addition of electron-hole ladder diagrams should considerably improve on the RPA's known inadequacy in
describing short-ranged correlations~\cite{OlsenRPA+ALDAEnergy}, and here we see that it affects both Hartree and
exchange-correlation parts.

\noindent$\bullet$
$\rho$ from QS\emph{GW} and $\Sigma$ from QS$G\widehat{W}$, which we denote as $\Sigma^{\mathrm{QS}G\widehat{W}}[\rho(\mathrm{QS}GW)]$
(bottom right panel of Fig.~\ref{fig:band}(f)).  This shows in a different way how the Hartree and correlation
contributions to the potential are interwined.

\begin{figure}
	\begin{center}
\includegraphics[width=0.5\columnwidth]{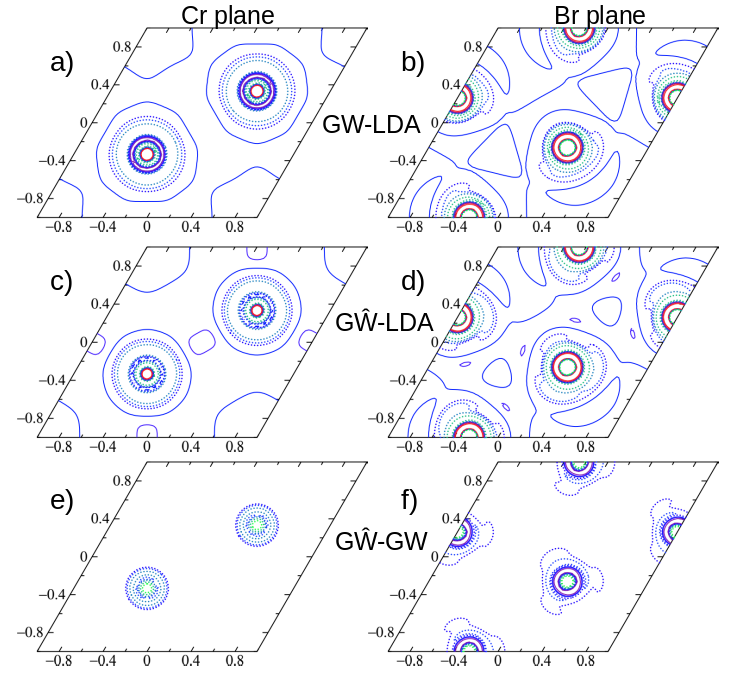}                         
		\caption{{\bf Charge densities ($\rho$) of \ce{CrBr3}}: of the highest valence band state 
			with the abscissa and ordinate \emph{x} and \emph{y}. 
			All the left panels (a,c,e) pass through a Cr plane and right panels (b,d,f) pass through a Br plane.  Panels (a,b)
			display constant-amplitude contours for QS\emph{GW} $\rho$ after subtracting out the LDA $\rho$. Contours are taken in half-decade
			increments in $\rho$, with a factor of 300 between highest contour (red) and lowest (blue).  Panels (c,d) show the change in
			$\rho$ passing from LDA to QS$G\widehat{W}$ eigenfunctions; panels (e,f) show the corresponding
			change passing from QS\emph{GW} to QS$G\widehat{W}$.  In the bottom four panels (c,d,e,f), blue{$\rightarrow$}red
			has a similar meaning as in the top panels (increasing positive ${\delta}\rho$), while contours
			of negative ${\delta}\rho$ are depicted by increasing strength in the change
			blue{$\rightarrow$}green.}
		\label{fig:charge}
	\end{center}
\end{figure}
To further probe the role of the charge density, we plot $\rho$ in the planes passing through the Cr and Br atoms at different
levels of theory (see Fig.~\ref{fig:charge}). The density is plotted in real space, and the abscissa and ordinate are defined by the the inverse transpose of the 2$\times$2 matrix composed of $\mathbf{b_1}$ and $\mathbf{b_2}$
(see Fig.~\ref{fig:struc}) with $x$ and $y$ defined by aligning $\mathbf{b_2}$ parallel to $y$.  In this notation the M point is on the $\mathbf{b_2}$ line, or the $y$ axis. On formation of the 2D crystal charge is augmented on the Cr-Cr and Br-Br
bonds, taking it away from the atoms.  QS\emph{GW} accentuates this tendency (see Fig.~\ref{fig:charge}(a,b)), as does QS$G\widehat{W}$, but to a relatively lesser extent (see Fig.~\ref{fig:charge}(c,d)). However, although the structure of the top most valence band seems similar within LDA and QS$G\widehat{W}$ and different within QS\emph{GW} and QS$G\widehat{W}$, we show in Fig.~\ref{fig:charge}(e,f) that the real-space $\rho(QSG\widehat{W})$ is much closer to $\rho(QSGW)$ compared to LDA.   In short, QS$G\widehat{W}$ weakly modifies and slightly localizes charges in comparison to QS\emph{GW}.


In conclusion, using a self-consistent first principles Green's function approach we show how correlations induce
large changes in both the one-body (Hartree) and many-body contribution to the potentials, and that the two are
inherently intertwined.  To demonstrate the effect we considered two currently popular materials systems: a
three dimensional charge-density-wave candidate TiSe$_{2}$ and a two-dimensional ferromagnet CrBr$_{3}$.  Such changes
to the electronic wavefunction go way beyond any weak renormalization of the parameters that determine the electronic
structure within a second principles approach and thus calls for development of better first principles approaches
that solve many-body Hamiltonians for real materials with better approximations.

MIK, ANR and SA are supported by the ERC Synergy Grant, project 854843 FASTCORR (Ultrafast dynamics of correlated
electrons in solids). MvS and DP are supported by the National Renewable Energy Laboratories.  We acknowledge PRACE for
awarding us access to Irene-Rome hosted by TGCC, France and Juwels Booster and Clusters, Germany, STFC Scientific
Computing Department's SCARF cluster.

\section*{method}
Single particle calculations (DFT, and energy band calculations with the static quasiparticlized QS\emph{GW} self-energy
	$\Sigma^{0}(k)$) were performed on a 16$\times$16$\times$16 (TiSe$_{2}$) and 16$\times$16$\times$1 (CrBr$_{3}$) \emph{k}-mesh while the (relatively smooth) dynamical self-energy
	$\Sigma(k)$ was constructed using a 8$\times$8$\times$8 (TiSe$_{2}$) and 6$\times$6$\times$1 (CrBr$_{3}$) \emph{k}-mesh and $\Sigma^{0}$(k) extracted from it.  For each
	iteration in the QS\emph{GW} self-consistency cycle, the charge density was made self-consistent.  The QS\emph{GW} cycle
	was iterated until the RMS change in $\Sigma^{0}$ reached 10$^{-5}$\,Ry.  Thus the calculation was self-consistent in
	both $\Sigma^{0}(k)$ and the density.  Numerous checks were made to verify that the self-consistent $\Sigma^{0}(k)$ was
	independent of starting point, for both QS$GW$ and QS$G\widehat{W}$ calculations; e.g. using LDA or Hartee-Fock self-energy as the initial self energy for QS\emph{GW} and using LDA or QS\emph{GW} as the initial
	self-energy	for QS$G\widehat{W}$.
	
	For the present work,  the electron-hole two-particle correlations are incorporated within a self-consistent ladder-BSE implementation~\cite{Cunningham18,cunningham21} with Tamm-Dancoff
	approximation~\cite{tamm1,tamm2}. The effective interaction \emph{W} is calculated with ladder-BSE corrections and the self energy, using a static vertex in the
	BSE.
	\emph{G} and \emph{W} are updated iteratively until all of them converge and this is what we call QS$G\widehat{W}$.  Ladders increase the screening of \emph{W},
	reducing the gap besides softening the LDA\textrightarrow{QS\emph{GW}} corrections noted for the valence bands. 

	For CrBr$_{3}$, we checked the convergence in the QS$G\widehat{W}$ band gap by increasing the size of the two-particle Hamiltonian. We
	increase the number of valence and conduction states that are included in the two-particle Hamiltonian. We observe that
	for all materials the QS$G\widehat{W}$ band gap stops changing once 24 valence and 24 conduction states are included in the
	two-particle Hamiltonian. While the gap is most sensitive to the number of valence states, 14 conducting states produces
	results within 2\% error of the converged results from 24 conduction states.

\section*{Competing interests}
The authors declare no competing financial or non-financial interests.
\section*{Correspondence}
All correspondence, code and data requests should be made to SA.
\section*{Data Availability}

All input/output data can be made available on reasonable request. All the input file structures and the commandlines to launch calculations are rigorously explained in the tutorials available on the Questaal webpage~\cite{questaal_web} \href{https://www.questaal.org/get/}.

\section*{Code Availability}
The source codes for LDA, QS\emph{GW} and QS$G\widehat{W}$ are available from~\cite{questaal_web}  \href{https://www.questaal.org/get/}  under the terms of the AGPLv3 license. 

\section*{Author contributions}
MIK and MvS conceived the main theme of the work.  SA, DP, MvS have carried out
the calculations. All authors have contributed to the writing of the paper and the analysis of the data.

\newpage
\pagebreak

%

\end{document}